\documentclass[AMA,STIX1COL]{WileyNJD-v2}
\usepackage{subfig}
\usepackage{graphicx}
\usepackage{tikz}
\usepackage{bm}
\usepackage{bigints}
\usepackage{multirow}

\usepackage{longtable}
\usetikzlibrary{shapes,decorations,arrows,calc,arrows.meta,fit,positioning}
\tikzset{
    -Latex,auto,node distance =1 cm and 1 cm,semithick,
    state/.style ={ellipse, draw, minimum width = 0.7 cm},
    point/.style = {circle, draw, inner sep=0.04cm,fill,node contents={}},
    bidirected/.style={Latex-Latex,dashed},
    el/.style = {inner sep=2pt, align=left, sloped}
}

\newcommand{\Esp}{\text{E}}

\newcommand{\pr}{\text{Pr}}

\articletype{Research article}%

\received{26 April 2016}
\revised{6 June 2016}
\accepted{6 June 2016}

\begin{document}

\title{~Longitudinal mediation analysis of time--to--event endpoints in the presence of competing risks}

\author[1,2]{Tat-Thang Vo*}

\author[3]{Hilary Davies-Kershaw}

\author[4]{Ruth Hackett}

\author[1,5]{Stijn Vansteelandt}

\authormark{Vo \textsc{et al}}

\address[1]{\orgdiv{Department of Applied Mathematics, Computer Science and Statistics}, \orgname{Ghent University}, \orgaddress{\state{Ghent}, \country{Belgium}}}

\address[2]{\orgdiv{Universit\'e de Paris}, \orgname{CRESS, INSERM, INRA}, \orgaddress{\state{Paris}, \country{France}}}

\address[3]{\orgdiv{Department of Population Health}, \orgname{London School of Hygiene and Tropical Medicine}, \orgaddress{\state{London}, \country{UK}}}

\address[4]{\orgdiv{Health Psychology Section, Department of Psychology}, \orgname{King's College London}, \orgaddress{\state{London}, \country{UK}}}

\address[5]{\orgdiv{Department of Medical Statistics}, \orgname{London School of Hygiene and Tropical Medicine}, \orgaddress{\state{London}, \country{UK}}}

\corres{*Tat-Thang Vo, Department of Applied Mathematics, Computer Science and Statistics, Ghent University, Krijgslaan 281, S9, Ghent 9000, Belgium. \email{TatThang.Vo@ugent.be}}

\abstract[Summary]{This proposal is motivated by an analysis of the English Longitudinal Study of Ageing (ELSA), which aims to investigate the role of loneliness in explaining the negative impact of hearing loss on dementia. The methodological challenges that complicate this mediation analysis include the use of a time-to-event endpoint subject to competing risks, as well as the presence of feedback relationships between the mediator and confounders that are both repeatedly measured over time. To account for these challenges, we introduce natural effect proportional (cause-specific) hazard models. These extend marginal structural proportional (cause-specific) hazard models to enable effect decomposition. We show that under certain causal assumptions, the path-specific direct and indirect effects indexing this model are identifiable from the observed data. We next propose an inverse probability weighting approach to estimate these effects. On the ELSA data, this approach reveals little evidence that the total efect of hearing loss on dementia is mediated through the feeling of loneliness, with a non-statistically significant indirect effect equal to 1.012 (hazard ratio (HR) scale; 95\% confidence interval (CI) 0.986 to 1.053).}


\jnlcitation{\cname{%
\author{Vo T.}, 
\author{Davies-Kershaw H.}, 
\author{Hackett R.}, and 
\author{Vansteelandt S.}} (\cyear{2020}), 
\ctitle{Longitudinal mediation analysis of time--to--event endpoints in the presence of competing risks}, \cjournal{Statistics in Medicine.}, \cvol{xxxx;xx:x--x}.}

\maketitle

\footnotetext{\textbf{Abbreviations:} ANA, anti-nuclear antibodies; APC, antigen-presenting cells; IRF, interferon regulatory factor}

\section{Introduction}\label{sec1}

This article is motivated by an analysis of the English Longitudinal Study of Ageing, a longitudinal cohort study of individuals aged 50 and older living in the community in England \citep{steptoe2013cohort, davies2017}, which follows participants biennially since 2002/03.
In previous studies, it was shown that both self-reported hearing loss and loneliness are significantly associated with a higher risk of dementia \citep{davies2017, davies18, rafnsson2020loneliness}. The question of interest, which we shall address here, is whether loneliness mediates the impact of hearing loss on incident physician-diagnosed dementia.

The focus on time-to-event endpoints (dementia) complicates the planned mediation analysis. It renders the popular difference- and product-of-coefficient methods inappropriate \citep{vdw16, robin92, pearl01}, necessitating the use of more complex causal mediation analysis methods. While such methods have been developed for the analysis of time-to-event endpoints, most ignore that the mediator is not assessed at baseline and that subjects may therefore experience the event prior to the mediator being assessed\citep{lange13,huang17,vandenberghe2018surrogate}. Further complications arise from the mediator, loneliness, being repeatedly measured. While useful to better capture mediation via the entire longitudinal mediator process \citep{vansteelandt19}, this also gives rise to complex time-varying confounding patterns whereby mediators (e.g. loneliness) and confounders (e.g. comorbidities) mutually influence each other over time. These complications have been addressed in a number of recent works \citep{zheng2017longitudinal,lin17,vansteelandt19}.

A further complication that we must consider, and that we have not previously found being addressed in the literature, is the presence of competing risks by death. With observed - as opposed to counterfactual - event times, the modelling of cause-specific hazards is well known to simplify the handling of competing risks. However, the modelling of (cause-specific) hazards is not readily possible in the previous mediation analysis works \citep{zheng2017longitudinal,lin17,vansteelandt19}, which instead focus on the analysis of survival chances. In view of this, in this paper, we will here introduce so-called natural effect proportional (cause-specific) hazard models, which directly parameterise the direct and indirect effects of a given exposure on the cause-specific hazard of the considered event. We will extend the weighting-based approach proposed by Mittinty and Vansteelandt \citep{mittinty2019longitudinal} for longitudinal natural effect models to enable the planned mediation analysis of an exposure via repeatedly measured mediators on a time-to-event endpoint subject to competing risks.

We proceed as follows. In the next section, we first describe the setting of interest. We then extend the natural effect proportional hazard model to take into account the longitudinal nature of the mediator. In the same section, we discuss the assumptions under which the path-specific direct and indirect effects derived from such model are identifiable from data, and provide a step-by-step procedure for estimating these effects. In section 3, we apply the proposed approach to analyze data from the English Longitudinal Study of Ageing \citep{steptoe2013cohort, davies2017}. We end with some final remarks and a discussion. 

\section{Proposal}\label{sec2}
\subsection{Natural Effect Models for longitudinal mediators and a time-to-event endpoint subject to competing risks}
Consider an observational study in which independent individuals $i = 1, \ldots, n$ are exposed to a categorical factor $A_i$ coded as $0, 1,\ldots, P-1$ for $P$ different categories. Longitudinal measurements of the mediator $M_{i0},M_{i1}\ldots, M_{iK}$ and of the covariates $L_{i0},L_{i1}\ldots, L_{iK}$ are subsequently recorded at baseline (subscript 0) and at visits $1, \ldots, K$, along with (i) a time-to-event endpoint $T_i$ and (ii) an index $D_i$ specifying whether the main event ($D_i = 1$) or the competing one ($D_i = 2$) happens. Denote $t_k$, $k = 1\ldots K$ the fixed time point after the onset of the exposure at which the measurements of $M_k$ and $L_k$ are pre-planned for all patients.  Assume that these measurements are only recorded until the last visit $K$ or until event $D_i = 1$ or $D_i = 2$ happens, whichever comes first. The time-to-event endpoint may be censored administratively or due to loss to follow-up, in which case $D_i = 0$.

The causal diagram in figure 1 depicts the relationships between the variables over time. It also represents a non-parametric structural equation model with independent errors. In the diagram, $L_k$ includes the indicator $I(T\ge t_k)$ of having survived visit $k$. Throughout, we will denote the history of measurements up to visit $k$ using a bar, i.e. $\overline{M}_k = (M_0, M_1,\ldots M_k)$.

To define the direct and indirect effects of interest, we will make use of so-called path-specific effects, expressed as a (cause-specific) hazard ratio. In particular, we define the counterfactual variables $T_{a,a^*}$ and $D_{a,a^*}$ as the time to the main or competing event (whichever comes first) and the corresponding event index that would be observed if the exposure $A$ were set to $a$ and the mediator levels changed to the levels that we would have seen if the exposure were set to $a^*$ and the levels of the time-varying confounders were as observed under this joint intervention on $A$ and $\overline{M}_K$, respectively. 
The total causal effect (TE) on the cause-$j$-specific hazard when the exposure changes from $a$ to $a^*$ is then expressed as
$ HR^j_{TE} = \frac{\lambda^j_{a^*, a^*}(t)}{\lambda^j_{a, a}(t)}$, which can be decomposed into the direct effect (DE) $HR^j_{DE} = \frac{\lambda^j_{a^*, a}(t)}{\lambda^j_{a, a}(t)}$ and the indirect effect (IE) $HR^j_{IE} = \frac{\lambda^j_{a^*, a^*}(t)}{\lambda^j_{a^*, a}(t)}$, where $HR^j_{TE} = HR^j_{IE}\times HR^j_{DE}$. The indirect effect $HR^j_{IE} $ hence reflects the part of the treatment effect (on the cause-$j$-specific hazard) that is mediated via the pathways $A \rightarrow M_k \rightarrow \ldots \rightarrow T$, where $k = 1, 2,\ldots$. These pathways start from the treatment $A$ and go directly to one of the mediator levels before getting to the event time $T$ by any intermediate path. In contrast, the direct effect $HR^j_{DE} $ reflects the part of the treatment effect (on the cause-$j$-specific hazard) that does not go through any of the above pathways.\\
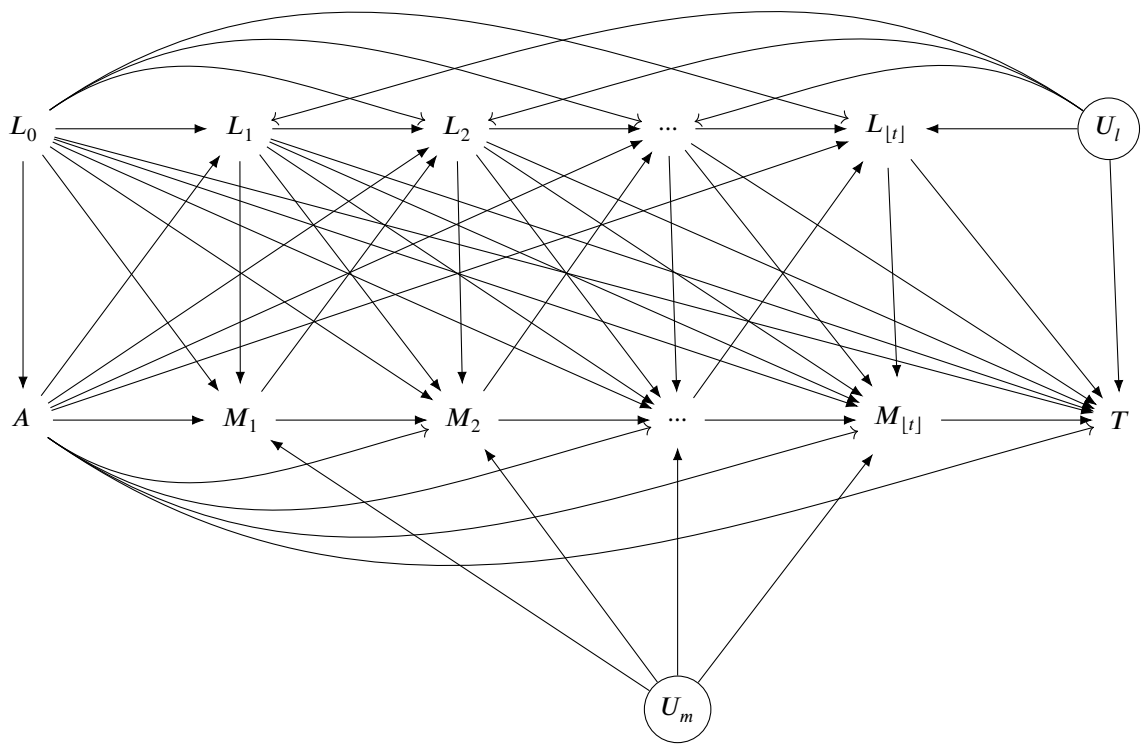
\begin{figure}
\centering
\scalebox{1}{\begin{tikzpicture}[node distance =3 cm and 2 cm]
    \node[state,draw = none] (a) at (0,0) {$A_{\,}$};

    \node[state,circle, draw = white] (m1) [right =of a] {$M_1$};
    \node[state,circle, draw = white] (m2) [right =of m1] {$M_2$};
    \node[state,circle, draw = white] (m3) [right =of m2] {$...$};
    \node[state,circle, draw = white] (mt) [right =of m3] {$M_{\lfloor t \rfloor}$};
    \node[state,circle, draw = white] (t) [right =of mt] {$T$};

    \node[state,circle, draw = white] (l0) [above =of a] {$L_0$};
    \node[state,circle, draw = white] (l1) [right =of l0] {$L_1$};
    \node[state,circle, draw = white] (l2) [right =of l1] {$L_2$};
    \node[state,circle, draw = white] (l3) [right =of l2] {$...$};
    \node[state,circle, draw = white] (lt) [right =of l3] {$L_{\lfloor t \rfloor}$};
    \node[state,circle](ul)[right =of lt]{$U_l$};
    \node[state,circle](um)[below =of m3]{$U_m$};

    \path (a) edge (m1); \path (m1) edge (m2);\path (m2) edge (m3);\path (m3) edge (mt);\path (mt) edge (t);
    \path (l0) edge (l1); \path (l1) edge (l2);\path (l2) edge (l3);\path (l3) edge (lt);\path (lt) edge (t);
    \path (l0) edge (a); \path (l0) edge (m1); \path (l0) edge (m2); \path (l0) edge (m3);\path (l0) edge (mt); \path (l0) edge (t);
    \path (a) edge (l1); \path (a) edge (l2);\path (a) edge (l3);\path (a) edge (lt);
    \draw [->] (a) to [out=325, in =195] (t);
    \draw [->] (a) to [out=325, in =195] (m2);
    \draw [->] (a) to [out=325, in =195] (m3);
    \draw [->] (a) to [out=325, in =195] (mt);
    \draw [->] (l0) to [out=35, in =165] (l2);
    \draw [->] (l0) to [out=35, in =165] (l3);
    \draw [->] (l0) to [out=35, in =165] (lt);
    \path (l1) edge (m1);\path (l1) edge (m2);\path (l1) edge (m3);\path (l1) edge (mt);\path (l1) edge (t);
    \path (l2) edge (m2);\path (l2) edge (m3);\path (l2) edge (mt);\path (l2) edge (t);
    \path (l3) edge (m3);\path (l3) edge (mt);\path (l3) edge (t);
    \path (lt) edge (mt);
    \path (m1) edge (l2);
    \path (m2) edge (l3);
    \path (m3) edge (lt);
    \path (ul) edge (lt);  \path (ul) edge (t);
    \draw [->] (ul) to [out=145, in =15] (l3);
    \draw [->] (ul) to [out=145, in =15] (l2);
    \draw [->] (ul) to [out=145, in =15] (l1);
    \path (um) edge (m1);\path (um) edge (m2);\path (um) edge (m3);\path (um) edge (mt);
\end{tikzpicture}}
\caption{Causal diagram.  $U_l$: unmeasured confounders affecting $L$ and $(T,D)$. $U_m$: unmeasured confounders affecting different measurements of $M$ over time}
\label{fig1}
\end{figure}

We are now ready to define the cause--$j$--specific natural effect proportional hazard model, accounting for longitudinal mediators, as follows:
 \begin{align}
\lambda^j_{a, a^*}(t) = \lambda_0^j(t)e^{\alpha_{1j}a + \alpha_{2j}a^*}
\end{align}
for all $a,a^*$, where $\lambda_0^j(t), \alpha_{1j}$ and $\alpha_{2j}$ are unknown. Under model $(1)$, the total, direct and indirect effect can be expressed on the hazard ratio scale as $HR^j_{TE} = e^{(\alpha_{1j}+\alpha_{2j})(a^* - a)}$; $HR^j_{IE} = e^{\alpha_{2j}(a^* - a)}$ and $HR^j_{DE} = e^{\alpha_{1j}(a^* - a)}$, 
respectively. To assess the possibility of mediator-exposure interaction, one can alternatively consider model:
 \begin{equation}\label{eq:model}
\lambda^j_{a, a^*}(t) = \lambda_0^j(t)e^{\alpha_{1j}a + \alpha_{2j}a^* + \alpha_{3j}a.a^*}
\end{equation}
Under model (2), the total causal effect on the hazard ratio scale is expressed as $HR^j_{TE} = e^{\left(\alpha_{1j}+\alpha_{2j} + \alpha_{3j}(a^*+a))\right)(a^* - a)}$ and is decomposed into the indirect effect $ HR^j_{IE} = e^{(\alpha_{2j} + \alpha_{3j}a)(a^* - a)}$ and the direct effect $HR^j_{DE} = e^{(\alpha_{1j} + \alpha_{3j}a^*)(a^* - a)}$. Finally, note that other models for survival outcomes, such as the Aalen model, can also be extended to the current context.
\subsection{Estimation}
To estimate the parameters in model $(1)$, we will assume that the set of baseline covariates $L_0$ is sufficient to control for confounding of the relationship between $A$ and $(T,D)$, as well as between $A$ and $M_t$ at any time. Besides, we assume there are no unmeasured confounders of the relationship between the time-to-event outcome and the mediator at any time. In figure 1, the latter assumption is satisfied since conditioning on the exposure $A$ and the history of the time-varying confounder $L$ up to time $t$ is sufficient to adjust for confounding of the relationship between the time-to-event outcome $T$ and the mediator level at time $t$. Our development allows for the presence of unmeasured common causes of the mediators over time (i.e. denoted $U_m$ in figure 1) and separate, independent unmeasured common causes of the baseline/time-varying confounders and the survival time $(T,D)$ (i.e. denoted $U_l$).

In what follows, we generalize the standard estimation procedure for the popular marginal strutural models to fit models (1) and (2). For this, we make the assumption that censoring is non-informative, in the sense that at any time, the instantaneous risk of the event among patients who then drop out of the study is not different (at all future times) from that of patients who remain, conditional on the exposure level. Under the aforementioned assumptions, Appendix 1 shows that consistent estimators of the parameters indexing model (1) can be obtained by solving the estimating equation:

\begingroup
\fontsize{10pt}{10pt}\selectfont
 \begin{align*}
\bigintsss_0^{\infty} \sum_{i, a, a^*}&\left\{\begin{pmatrix} a \\ a^* \end{pmatrix} - \frac{\sum_{a,a^*}\hat E\left[\begin{pmatrix} a \\ a^* \end{pmatrix}
R_i^j(t)W_i(\lfloor t \rfloor ,a,a^*)e^{\alpha_{1j}a + \alpha_{2j}a^*}\right]}
{\sum_{a,a^*} \hat E\left[R_i^j(t)W_i(\lfloor t \rfloor ,a,a^*)e^{\alpha_{1j}a + \alpha_{2j}a^*}\right]}\right\}\cdot \\
& \cdot R_i^j(t)W_i(\lfloor t \rfloor ,a,a^*)\left(dN_i^j(t) - \lambda_0^j(t)e^{\alpha_{1j}a + \alpha_{2j}a^*}\right) = 0
\end{align*}
\endgroup
where $R_i^j(t) = I(T_i\ge t, D_i = j)$;  $dN^j_i(t) = I(T_i = t, D_i = j)$ and $I(.)$ denotes the indicator function. Besides,
\begin{align*}
W_i(\lfloor t \rfloor ,a,a^*) = \frac{\prod_{s:t_s\le \lfloor t \rfloor } \pr(M_{s,i}|A_i = a^*, \overline{M}_{s - 1,i}, \overline{L}_{s,i}, T_i \ge t_s)}{\prod_{s:t_s\le \lfloor t \rfloor} \pr(M_{s,i}|A_i = a, \overline{M}_{s - 1,i}, \overline{L}_{s,i}, T_i \ge t_s)}\times \frac{I(A_i = a)}{\pr(A_i=a|L_{0,i})}
\end{align*}
denotes the weight of individual $i$ at time $t$ and at exposure levels $a$ and $a^*$. The second component of the weight ensures that the exposure-outcome association is adjusted for confounding by $L_0$. It creates a pseudo-population in which the exposure is no longer associated with $L_0$ and hence removes confounding by $L_0$ \citep{lange12}. The first component of the weight then distinguishes between the direct and indirect paths by correcting for the fact that the observed mediator value at each time point before $t$ may differ from the counterfactual value that is of interest at that time. Note that the notation $\lfloor t \rfloor$ here is slightly different from its standard definition, to take into account the fact that if a patient experiences an event or leaves the study at time $t=t_k$ of visit $k$, no measurement of $M_k$ and $L_k$ is possible at that time. More precisely,
\[\lfloor t \rfloor= \begin{cases}
t_{k-1}~~~\mathrm{if}~~~ t=t_k\\
t_k~~~~~~~\mathrm{if}~~~ t_k < t < t_{k+1}
\end{cases}\]
where $ k =1,\ldots,K$. From this, the fitting procedure is described as follows:

\textbf{Step 1--} Postulate and fit a suitable model for the exposure $A$ conditional on the baseline confounders ($L_0$) based on the original data set. For instance, a multinomial logistic model can be used for a categorical exposure with $P$ possible values:
\begin{align}
\log \frac{\pr (A = a|L_0 = l_0)}{\pr (A = 0|L_0 = l_0)} = \beta_{0,a} + \beta_{1,a}\,l_0,
\end{align}
where $a = 1, \ldots, P - 1$ and $\pr (A = 0|L_0 = l_0) = 1/(1 + \sum_{a = 1}^{P - 1} e^{\beta_{0,a} + \beta_{1,a}\,l_0})$.

\textbf{Step 2--}  Convert the original dataset to a long or counting-process format, in which the observation period $[0, t_i]$ of subject $i$ is broken into $\lfloor t_i \rfloor + 1$ intervals if $t_i>\lfloor t_i \rfloor>t_1$, into $t_i$ intervals if $t_i=\lfloor t_i \rfloor >t_1$ and is kept unchanged if $\lfloor t_i \rfloor<t_1$. Each interval $k$ will have the following information encoded:
\begin{enumerate} 
\item[(a)] The beginning of the interval, which equals the time $t_{k-1}$ of visit $k-1$, with $t_0 = 0$.
\item[(b)] The end of the interval, which equals the time $t_k$ of visit $k$ or the event/censoring time $t_i$ for the last interval.
\item [(c)] The event status at the end of the interval.
\item[(d)] The exposure $A_i$ and the baseline covariates $L_{0i}$, whose values remain unchanged across all intervals.
\item[(e)] The history of the mediator and the longitudinal confounders recorded up to the end of the interval. Note that the history of $L$ and $M$ for subject $i$ up to the time $t_i$ is similar to their history up to the last visit prior to time $t_i$. 
\end{enumerate}
Table 1 provides a toy example in which three patients receive a binary treatment and are followed up for a total of three years, with two visits pre-planned at the end of year 1 and 2. Patient 3 is free of event till the end of the study and hence has the mediator level fully recorded at the two intermediate visits. In contrast, patient 1 and 2 experience an event after 1.5 and 0.9 years, due to which they have no (i.e. patient 2) or only one (i.e. patient 1) mediator level recorded. Table 2 illustrate how the information of these three hypothetical individuals is encoded in a counting-process format. 

\begin{table}
\centering
\caption{Illustrating example: A toy dataset in standard short format. Here, $m_{ij}$ and $l_{ij}$ denote the mediator and longitudinal confounder level of individual $j$ recorded at visit $t_i$, respectively. The treatment $A$ is binary (0 vs. 1) and there are two competing events, coded as $1$ and $2$.} 
\label{tab:ma}
\begin{tabular}{c c c c c c c c c}
\hline
   Individual & Following-up time  & Status & $A$ & $M_1$ & $M_2$ & $L_0$ & $L_1$ & $L_2$\\
&(years)\\ 
\hline
1 & 1.5 & 1 & 1 & $m_{11}$ & $-$      & $l_{01}$ & $l_{11}$ & $-$  \\
2 & 0.9 & 2 & 0 &  $-$   & $-$      & $l_{02}$ & $-$ & $-$ \\
3 & 3.0 & 0 & 1 & $m_{13}$ & $m_{23}$  & $l_{03}$ & $l_{13}$  & $l_{23}$\\
\hline
\end{tabular}
\end{table}
\begin{table}
\centering
\caption{The counting-process format of the dataset in table 1} 
\label{tab:ma}
\begin{tabular}{c c c c c c c c c c}
\hline
   Individual & Start & Stop & Status & $A$ & $M_t$ & $M_{t-1}$ & $L_t$ & $L_{t-1}$ & $L_0$\\ 
\hline
1 & 0 & 1    & 0 & 1 & 0   & 0     & 0     & 0    & $l_{01}$\\
1 & 1 & 1.5 & 1 & 1 & $m_{11}$ & 0     & $l_{01}$ & 0 & $l_{01}$ \\
2 & 0 & 0.9 & 2 & 0 & 0    & 0     & 0     & 0    & $l_{02}$\\
3 & 0 &    1 & 0 & 1 & 0    & 0     & 0     & 0    & $l_{03}$\\
3 & 1 &    2 & 0 & 1 & $m_{13}$ & 0 & $l_{13}$  & 0 & $l_{03}$ \\
3 & 2 &    3 & 0 & 1 & $m_{23}$ & $m_{13}$ & $l_{23}$ & $l_{13}$  & $l_{03}$\\
\hline
\end{tabular}
\end{table}

\textbf{Step 3--}  Postulate and fit a suitable model for the mediator at a time point $t$, conditional on the exposure, the longitudinal confounder $\overline{L}_t$ and previous measurements of the mediator (i.e. $\overline{M}_{t - 1}$), by using the long data set. For instance, one may assume multinomial logistic models for a categorical mediator $M_k$ with possible values $0, \ldots, Q$. The model for $M_k$ is thus:
\begin{align}
\log \frac{\pr(M_k = q|A = a, \overline{M}_{k - 1} = \overline{m}_{k - 1}, \overline{L}_k = \bar{l}_k, T\ge t_k)}{\pr(M_k = 0|A = a, \overline{M}_{k - 1} = \overline{m}_{k - 1}, \overline{L}_k =\overline{l}_k, T\ge t_k)} = \gamma_{0q} + \gamma_{1q} a + \gamma_{2q}'\overline{m}_{k - 1} + \gamma_{3q}'\overline{l}_t
\end{align}
where $q = 1, \ldots, Q$. 

\textbf{Step 4--}  A new data set is then constructed by copying the original data set (in long format) $P$ times and including an additional variable $A^*$ to capture the $P$ possible values of the exposure relative to the indirect path. $A^*$ is set to the actual value of the exposure $A$ for the first replication, to the other potential values of $A$ for the remaining replications. For the example discussed in table 1 and 2, the corresponding extended data set is provided in table 3.
\begin{table}
\centering
\caption{ Extended data set for the example in Table 1} 
\label{tab:ma}
\begin{tabular}{c c c c c c c c c c c}
\hline
   Individual & Start & Stop & Status & $A$ & $A^*$ & $M_t$ & $M_{t-1}$ & $L_t$ & $L_{t-1}$ & $L_0$\\ 
\hline
1 & 0 & 1    & 0 & 1 & 1 & 0   & 0     & 0     & 0    & $l_{01}$\\
1 & 1 & 1.5 & 1 & 1 & 1 &  $m_{11}$ & 0     & $l_{01}$ & 0 & $l_{01}$ \\
2 & 0 & 0.9 & 2 & 0 & 0 &  0    & 0     & 0     & 0    & $l_{02}$\\
3 & 0 &    1 & 0 & 1 & 1 &  0    & 0     & 0     & 0    & $l_{03}$\\
3 & 1 &    2 & 0 & 1 & 1 &  $m_{13}$ & 0 & $l_{13}$  & 0 & $l_{03}$ \\
3 & 2 &    3 & 0 & 1 & 1 & $m_{23}$ & $m_{13}$ & $l_{23}$ & $l_{13}$  & $l_{03}$\\
\hline
1 & 0 & 1    & 0 & 1 & 0 & 0   & 0     & 0     & 0    & $l_{01}$\\
1 & 1 & 1.5 & 1 & 1 & 0 &  $m_{11}$ & 0     & $l_{01}$ & 0 & $l_{01}$ \\
2 & 0 & 0.9 & 2 & 0 & 1 &  0    & 0     & 0     & 0    & $l_{02}$\\
3 & 0 &    1 & 0 & 1 & 0 &  0    & 0     & 0     & 0    & $l_{03}$\\
3 & 1 &    2 & 0 & 1 & 0 &  $m_{13}$ & 0 & $l_{13}$  & 0 & $l_{03}$ \\
3 & 2 &    3 & 0 & 1 & 0 & $m_{23}$ & $m_{13}$ & $l_{23}$ & $l_{13}$  & $l_{03}$\\
\hline
\end{tabular}
\end{table}

\textbf{Step 5--}  Compute weights by applying the fitted models from steps 1 and 3 to the new data set. At visit $k$, the weight for the $i^{th}$ individual is $w_{i}(k, a,a^*) = w_{i}^{ttm}(a)\cdot w_{i}^{med}(k,a,a^*),$ where:
\[w_i^{ttm}(a) = \sum_{a = 0}^{P-1} \frac{I(A_i = a)}{\pr(A_i = a|L_{0i} = l_{0i})}\]
and
\[w_{i}^{med}(k,a,a^*) = \prod_{s: t_s \le t_k} \frac{\pr(M_{s,i} = m_{s,i}|A_i = a^*, \overline{M}_{s - 1,i} = \overline{m}_{s - 1,i}, \overline{L}_{s,i} = \bar{l}_{s,i}, T_i\ge t_s)}{\pr(M_{s,i} = m_{s,i}|A_i = a, \overline{M}_{s - 1,i} = \overline{m}_{s - 1,i}, \overline{L}_{s,i} =\overline{l}_{s,i}, T_i\ge t_s)}\]
where the subscript $ttm$ and $med$ denotes treatment and mediator, respectively. At the end of the follow-up time, the weight for a patient having $T_i = t_i$ is $w_{i}(t_i, a,a^*) = w_{i}(\lfloor t_i \rfloor ,a,a^*)$. 

\textbf{Step 6--}  Fit the natural effect cause-specific proportional hazard model (1) and (2) by proportiona hazard regression of the cause-specific event time on $A$ and $A^*$ on the basis of the expanded data set, using the weights computed in the previous step. 

\textbf{Step 7--}  Derive confidence intervals for the parameters in model (1) and (2) by using the non-parametric bootstrap. For this, one first generates $S$ bootstrap samples with replacement from the original dataset, then repeats all the above steps for each bootstrap sample. The 95\% confidence interval for each parameter in model (1) and (2) is computed by using the 2.5\% and 97.5\% quantiles of the bootstrap distribution of the corresponding estimator.

\subsection{Addressing complications due to censoring}
Denote $C$ the time-to-censoring. As stated above, when the censoring is non-informative conditional on the exposure (figure 2a), the provided estimating procedure remains valid without further adjustment. When censoring is dependent upon the baseline covariate vector $L_0$ and the exposure $A$, one could adjust for censoring by alternatively focusing on the so-called conditional cause-specific natural effect proportional hazard model, that is,
\begin{align}
\lambda^j_{a, a^*}(t|L_0 = l_0) = \lambda_0^j(t)e^{\alpha_{1j}a + \alpha_{2j}a^* + \alpha_{2j}'l_0}
\end{align}
Note that an interaction between $a^*$ and $l_0$ could also be permitted in such model to assess the possibility of mediator-baseline covariate interaction. The procedure discussed in section 2.2 can then be applied to estimate the parameters in this model, with a slight adjustment in step 6 where apart from $A$ and $A^*$, the covariates $L_0$ (and the product of $A^*$ and $L_0$ if mediator- baseline covariate interaction is assessed) are also included into the proportional hazard regression model.
\begin{figure}
\begin{center}
\subfloat[][]{\begin{tikzpicture}[node distance =1.5 cm and 1.5 cm]
    \node[state,draw = none] (a) at (0,0) {$A_{\,}$};

    \node[state,circle, draw = white] (mt) [right =of a] {$\overline{M}_t$};
    \node[state,circle, draw = white] (t) [right =of mt] {$T$};

    \node[state,circle, draw = white] (l0) [above =of a] {$L_0$};
    \node[state,circle, draw = white] (lt) [right =of l0] {$\overline{L}_t$};
    \node[state,circle, draw = white] (c) [below =of a] {$R_C(t)$};

    \node[state,circle](ul)[right =of lt]{$U_l$};
    \node[state,circle](um)[below =of mt]{$U_m$};

    \path (a) edge (mt);\path (mt) edge (t);
    \path (l0) edge (lt); \path (lt) edge (t);
    \path (l0) edge (a); \path (l0) edge (mt); \path (l0) edge (t);
    \path (a) edge (lt);
    \path (a) edge (c);
    \draw [->] (a) to [out=325, in =215] (t);
    \path (lt) edge (mt);
    \path (ul) edge (lt);  \path (ul) edge (t);
    \path (um) edge (mt);
\end{tikzpicture}}
\hspace{10pt}
\subfloat[][]{\begin{tikzpicture}[node distance =1.5 cm and 1.5 cm]
    \node[state,draw = none] (a) at (0,0) {$A_{\,}$};

    \node[state,circle, draw = white] (mt1) [right =of a] {$\overline{M}_{t-1}$};
    \node[state,circle, draw = white] (mt) [right =of mt1] {$M_t$};
    \node[state,circle, draw = white] (t) [right =of mt] {$T$};

    \node[state,circle, draw = white] (l0) [above =of a] {$L_0$};
    \node[state,circle, draw = white] (lt1) [right =of l0] {$\overline{L}_{t-1}$};
    \node[state,circle, draw = white] (lt) [right =of lt1] {$\overline{L}_t$};
    \node[state,circle, draw = white] (c) [below =of mt1] {$R_C(t)$};

    \node[state,circle](ul)[right =of lt]{$U_l$};
    \node[state,circle](um)[below =of mt]{$U_m$};

    \path (a) edge (mt1);\path (mt1) edge (mt);\path (mt) edge (t);
    \path (l0) edge (lt1); \path (lt1) edge (lt); \path (lt) edge (t);
    \path (l0) edge (a); \path (l0) edge (mt1); \path (l0) edge (mt); \path (l0) edge (t);
    \path (lt1) edge (mt1);
    \path (lt1) edge (mt);
    \path (lt1) edge (t);
    \path (mt1) edge (c);
    \path (mt1) edge (lt);

    \path (a) edge (lt);
    \path (a) edge (lt1);
    \path (a) edge (c);
    \draw [->] (a) to [out=325, in =215] (t);
    \path (lt) edge (mt);
    \path (ul) edge (lt);  \path (ul) edge (t);
    \path (um) edge (mt); \path (um) edge (mt1);
    
     \draw [->] (lt1) to [out=225, in =115] (c);
\end{tikzpicture}}
\end{center}
\caption{(Simplified) causal diagram when censoring presents -- (a) Censoring is non-informative conditional on the exposure and (b) Censoring is non-informative at time $t$ conditional on the exposure and the history up to that time}
\end{figure}
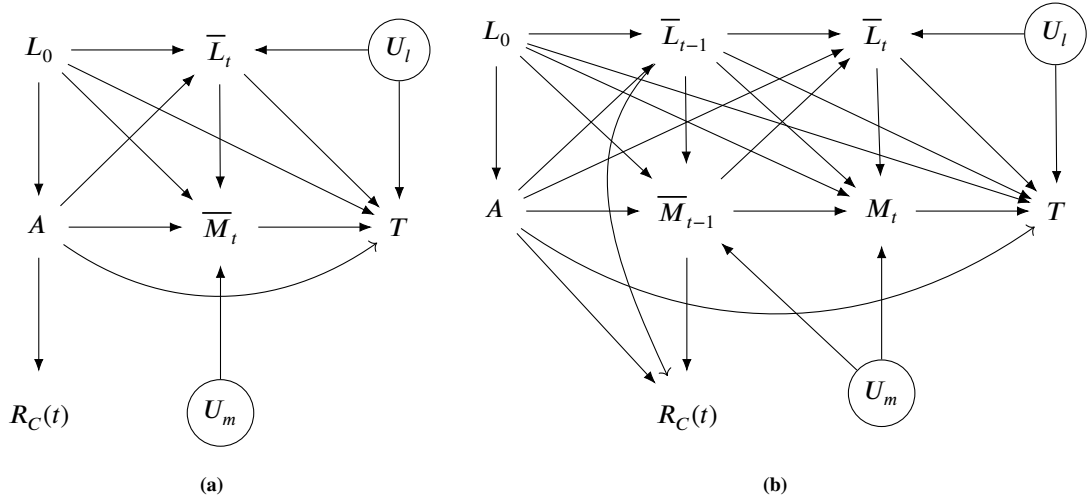
In practice, it might however be the case that censoring is dependent upon post-baseline factors such as the longitudinal mediator and confounder levels that are measured prior to censoring (figure 2b). In Appendix A2, we show that if at any time $t$, the risk of future events for patients who drop out of the study is not different from that of patients who have the same exposure, mediator and covariate history up to time $t$ but remain in the study, the so-called inverse probability of censoring weighting approach can be used to account for censoring. More precisely, the parameters indexing models (1) and (2) are then estimated by solving the following equation:
 \begin{align*}
&\bigintssss_0^\infty \sum_{i, a, a^*}\left\{\begin{pmatrix} a \\ a^* \end{pmatrix} - \frac{\sum_{a,a^*}\hat E\left[\begin{pmatrix} a \\ a^* \end{pmatrix}\cdot R_i(t)\cdot I(C_i>t)\cdot W_i(\lfloor t \rfloor )\cdot e^{\alpha_{1j}a + \alpha_{2j}a^*}\right]}
{\sum_{a,a^*} \hat E\left[R_i(t)\cdot I(C_i>t)\cdot W_i(\lfloor t \rfloor )\cdot e^{\alpha_{1j}a + \alpha_{2j}a^*}\right]}\right\}\cdot \\
&\,\,\,\,\,\,\,\,\,\,\,\,\,\,\,\,\,\,\,\,\,\,\,\,\,\,\,\,\,\,\cdot R_i(t)\cdot I(C_i>t)\cdot W_i(\lfloor t \rfloor )\left(dN_i(t) - \lambda_0(t)\cdot e^{\alpha_{1j}a + \alpha_{2j}a^*}dt\right) = 0
\end{align*}
where
\begin{align*}
W_i(\lfloor t \rfloor,a,a^*) =& \frac{\prod_{s: t_s \le \lfloor t \rfloor} \hat\pr(M_{s,i}|A_i = a^*, \overline{M}_{s-1,i}, \overline{L}_{s,i}, T_i\ge t_s, C_i \ge t_s)}{\prod_{s: t_s \le \lfloor t \rfloor} \hat\pr(M_{s,i}|A_i = a, \overline{M}_{s-1,i}, \overline{L}_{s,i}, T_i\ge t_s, C_i \ge t_s)}\cdot \frac{I(A_i = a)}{\hat\pr(A_i=a|L_{0,i})}\cdot\\
&\,\cdot\frac{1}
{\bm\prod_{s: 0\le s \le t} \left[1- \hat{\lambda}_C(s|T_{i}>s, \overline{M}_{\lfloor s\rfloor ,i}, \overline{L}_{\lfloor s\rfloor,i},L_{0,i},A_i=a)\right]}
\end{align*}
and $\lambda_C(.)$ denoting the cause-specific hazard function of the time-to-censoring. Here, $\bm\prod_{s} x_s$ is defined as a product limit. With the above weight $W_i(\lfloor t \rfloor,a,a^*)$, the additional component that accounts for the informative censoring can make the overall weight become unstable (e.g. when the censoring hazard is close to 1 in some
strata). To overcome this, one can then use stabilized (censoring) weights which incorporate a numerator defined in the same way as the denominator but adjusting only for the exposure, that is:
\[\frac{\bm\prod_{s: 0\le s \le t} \left[1- \hat{\lambda}_C(s|A_i=a)\right]}
{\bm\prod_{s: 0\le s \le t} \left[1- \hat{\lambda}_C(s|T_{i}>s, \overline{M}_{\lfloor s\rfloor ,i}, \overline{L}_{\lfloor s\rfloor,i},L_{0,i},A_i=a)\right]}\]
One then needs to postulate two models for the censoring hazard at time $t$, with one conditioning on exposure and the other conditioning on exposure, baseline covariates and the history of the longitudinal mediator and confounders up to time $t$, where only the latter model needs to be correct. For instance, $\lambda_C(t|A_i = a_i)=\lambda'_{0C}(t)e^{\eta a_i}$ and 
\begin{align} \lambda_C(t|T_{i}>t, \overline{M}_{\lfloor s\rfloor,i}=\overline{m}_{\lfloor s\rfloor,i}, \overline{L}_{\lfloor s\rfloor,i}=\overline{l}_{\lfloor s\rfloor,i},L_{0,i}=l_{0,i},A_i=a_i)=\lambda_{0C}(t)e^{\theta_0a_i + \theta_1l_{0,i} + \theta_2'\overline{m}_{\lfloor s\rfloor ,i} + \theta_3'\overline{l}_{\lfloor s\rfloor,i}}\end{align} 
As a result, in step 5 of the estimation procedure, apart from computing the mediator and treatment weights, one needs to additionally derive the censoring weight. More precisely, the weight for the $i^{th}$ individual at visit $k$ is now $w_i^{ttm}(a,a^*)\cdot w_i^{med}(k,a)\cdot w_i^{cen}(k,a)$, where the subscript \textit{cen} denotes censoring, i.e.,
\[w_i^{cen}(k,a)=\frac{1}{\prod_{s: 0\le s \le t_k} \left[1- \hat{\lambda}_C(s|T_{i}>s, \overline{m}_{s ,c,i}, \overline{l}_{s,c,i},l_{0,i},a)\right]}\]
 while $w_i^{ttm}(a,a^*)$ and $w_i^{med}(k,a,a^*)$ are computed as above. If a proportional hazard censoring model as (6) is fitted, the baseline censoring hazard $\lambda_{0C}(t)$ and $\lambda'_{0C}(t)$ at time $t$ can be estimated by the standard Breslow estimator. Once the individual weights are computed, the natural effect cause-specific proprotional hazard model (1) and (2) can be fitted using these weights and the confidence intervals for the estimates can be derived via the nonparametric bootstrap, as described above.
\section{Illustrating example}
 We illustrate the proposed approach on the ELSA data. In this ongoing study, the first contact with the participants was in 2002/03 (wave 1). These participants were then followed up biennially, with measures collected via computer-assisted face-to-face interview and self-completion questionnaires. For this illustration, the data are available until one year after the last wave in 2016/17. As stated above, the question of interest here is whether the feeling of loneliness mediates the impact of hearing loss on dementia, accounting for mortality as a competing event. 

For this analysis, we used the hearing measurement recorded at wave 2 (e.g. 2004/05) and dichotomized subjects into two groups, namely normal ($A=0$) and limited ($A=1$) hearing ability. The longitudinal mediator "loneliness" was recorded from wave 3 (2006/07) to wave 7 (2014/15) and had two potential values, namely frequent ($M=1$) vs. infrequent feeling of loneliness ($M=0$). Alongside the mediator, four longitudinal confounders were recorded over time (i.e. wave 3 to wave 7), namely depression status (yes vs. no), mobility score (continuous), smoking status (non-smoker vs. current smoker) and alcohol status (non-drinker vs. current drinker). The baseline covariates consisted of age, gender, ethnicity (white vs. non-white), wealth (1=low, 5=high), education level (1 = no formal qualification, 2 = intermediate and 3 = higher education), marital status (yes vs. no), the use of hearing aids (yes vs. no), the presence of other comorbidities (i.e. hypertension, diabetes, stroke and cancer -- yes vs. no) and the baseline values of the aforementioned time-varying confounders. A detailed description of these covariates was provided elsewhere  \citep{davies2017, hackett2018walking, davies18, rafnsson2020loneliness}.

We assumed that the relationship between the variables obeys the causal structure depicted in figure 1. Here, the mediators and confounders measured at time $t_{k - 1}$ are time-varying confounders of the relationship between the mediator measured at time $t_k$ and the outcome. To derive the treatment weights (step 1), we first considered a logistic (treatment) model adjusting for the main effects of all baseline covariates. To assess the potential of covariate-covariate interactions, we used a LASSO variable selection process (R package glmnet) to select the most important interaction terms from the set of all possible two-by-two covariate interactions. The chosen interactions were then added into the treatment model. The tuning parameter in the LASSO was selected by leave-one-out cross-validation. 

To derive the mediation weights (step 3), we first considered a logistic (mediator) model adjusting for $t_k$, $M_{k-1}$, $L_k$ and the main effects of all baseline covariates. To assess whether the conditional distribution of $M_k$ had a residual dependence upon the history of $M$ and $L$ that preceded the time $t_{k-1}$ (for $M$) and $t_k$ (for $L$), we used the LASSO to determine the first post-baseline measurement of $M$ and $L$ that were predictive for $M_k$, conditional on the later measurements. This measurement and all measurements following this one were included into the mediator model. Next, we assessed whether there were important (i) treatment-baseline/longitudinal covariate interactions, (ii) time-baseline covariate interactions and (iii) baseline covariate-covariate interactions that should be adjusted for. For each step, an independent LASSO variable selection process was performed to select the most important interaction terms from all possible interactions of the same type. The interactions that were chosen in the previous step were always included in the model of the subsequent steps (which implies no shrinkage on these terms in the subsequent steps). The tuning parameter in each LASSO procedure was selected by leave-one-out cross-validation. The final model was refitted before calculating the mediation weights.

To derive the censoring weights, we first considered a cause-specific proportional hazard model adjusting for $M_{k}$, $L_k$, the exposure $A$ and the main effects of all baseline covariates $L_0$. To assess whether the censoring hazard at time $t$ had a residual dependence upon the history of $M$ and $L$ that preceded the time $t_{k}$ for $M$ and for $L$, we implemented a backward elimination process, using the Akaike information criterion to determine the first post-baseline measurement of $M$ and $L$ that were predictive for the censoring hazard at time $t$, conditional on the later measurements. This measurement and all measurements following this one were included into the censoring proportional hazard model. Next, as for the mediator model, we assessed whether there were important (i) treatment-baseline/longitudinal covariate interactions and (ii) baseline covariate-covariate interactions that should be adjusted for. For each step, an independent backward elimination process was performed to select the most important interaction terms from all possible interactions of the same type. The interactions that were chosen in the previous step were always included in the model of the subsequent steps (which implies no exclusion of these terms in the subsequent steps). Note that we used backward elimination for the construction of the censoring models (as opposed to LASSO) due to the lack of prepackaged software that can apply LASSO or other penalized variable selection methods on a counting format survival dataset. Results of the variable selection processes for the treatment, mediator and censoring models are reported in Online Supplementary Material file.

The two natural effect proportional hazard models (i.e. model (1) and (2)) specific for dementia and for death were then fitted using the calculated weights. The confidence intervals of the total, direct and indirect hazard ratios were derived by the non-parametric bootstrap method, with 5000 samples taken from the original data set by sampling with replacement. We then established the cumulative incidence curves of dementia and of death under different sets of $a$ and $a^*$. These curves reflect the cumulative failure rates over time for a particular cause (e.g. dementia), acounting for the presence of other competing events (e.g. death). To estimate the curves, we considered dementia and death as two terminal states of a multi-state model where the transition from dementia to death was treated as an absorbing state, i.e. the one that subjects never exist \citep{putter07}. For pedagogic purposes, we only use the results of the natural effect model (1) (i.e. without interaction between $a$ and $a^*$) to establish these curves. 
\begin{figure}
\centering\includegraphics[width = \textwidth, height = \textheight,keepaspectratio]{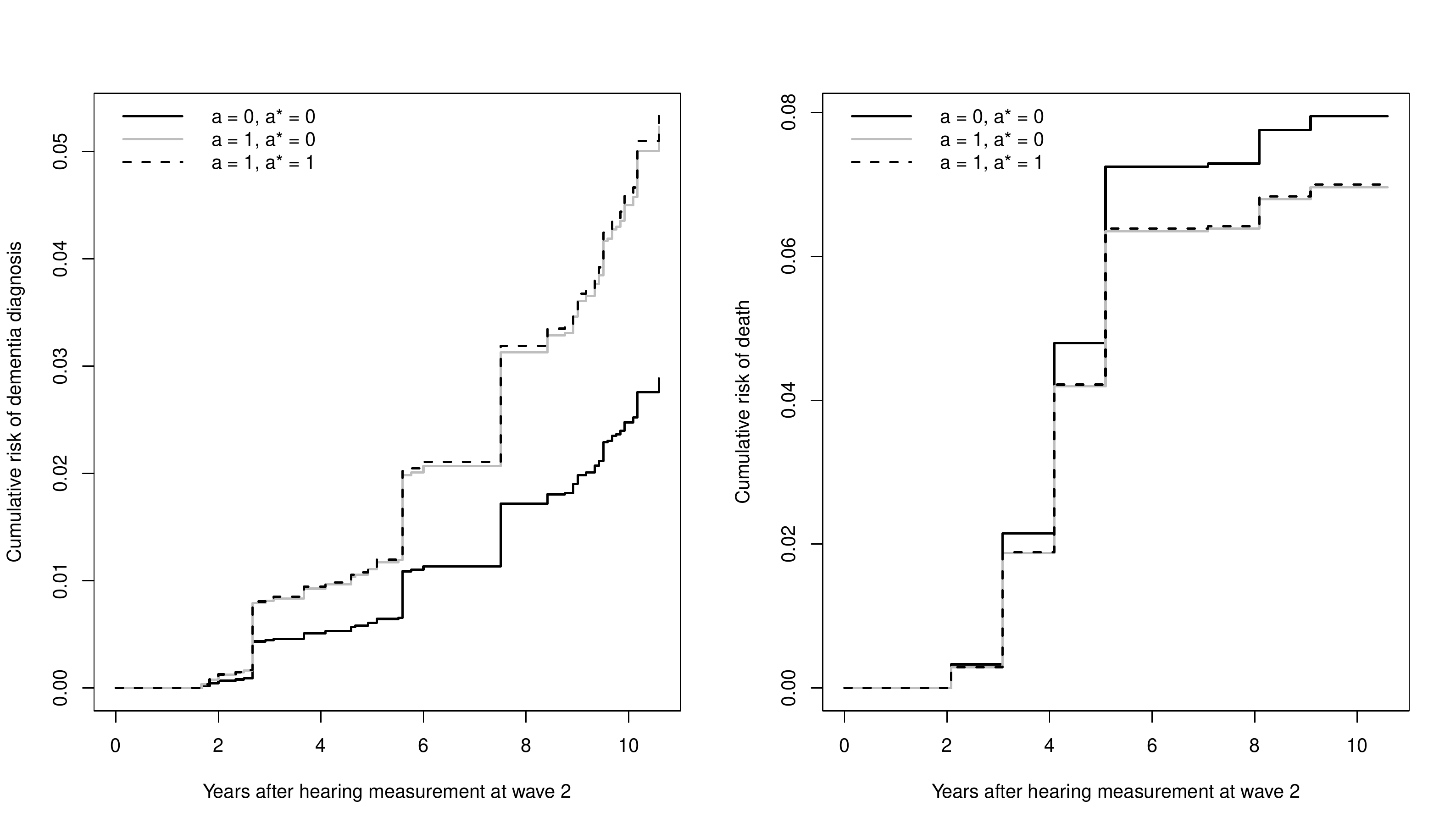}
\caption{Estimated cumulative incidence curve of dementia diagnosis}
\label{fig2}
\end{figure}

\begin{table}
\centering
\caption{Data analysis: estimation of the natural effect models} 
\label{tab:ma}
\begin{tabular}{c l c c c c}
\hline
Model & Coefficient & Estimate & 95\%CI & p-value\\ 
\hline
\multirow{6}{*}{(1)} & \textbf{Dementia} ($j=1$) & & &  \\
 &\quad$\alpha_{1j}$ & 0.599 & (0.090; 1.001) & 0.007\\
 &\quad$\alpha_{2j}$ & 0.012 & (-0.014; 0.051) & 0.438\\
  & \textbf{Death} ($j=2$) & & &  \\
 &\quad$\alpha_{1j}$ & -0.126 & (-0.483; 0.143) & 0.462\\
 &\quad$\alpha_{2j}$ & 0.001 & (-0.013; 0.011) & 0.913\\
 \hline
\multirow{8}{*}{(2)} & \textbf{Dementia} ($j=1$) & & &  \\
 &\quad$\alpha_{1j}$ & 0.591 & (0.093; 1.003) & 0.007\\
 &\quad$\alpha_{2j}$ & 0.002 & (-0.012; 0.017) & 0.827\\
 &\quad$\alpha_{3j}$ & 0.017 & (-0.023; 0.079) & 0.520\\
  & \textbf{Death} ($j=2$) & & &  \\
 &\quad$\alpha_1$ & -0.121 & (-0.480; 0.148) & 0.485\\
 &\quad$\alpha_2$ & 0.005 & (-0.011; 0.023) & 0.584\\
 &\quad$\alpha_3$ & -0.009 & (-0.049; 0.022) & 0.632\\
\hline
\end{tabular}
\end{table}

\begin{table}
\centering
\caption{Data analysis: the effect of limited vs. normal hearing ability on the time-to-event outcomes, mediated through the feeling of loneliness. The mediated proportion is calculated on log scale} 
\label{tab:ma}
\begin{tabular}{c l c c c}
\hline
Model & Effect & Hazard ratio & 95\%CI & Mediated proportion\\ 
\hline
\multirow{8}{*}{(1)} & \textbf{Dementia} & & &  \\
 &Total effect  ($A=1$ vs. $A=0$)  & 1.843 & (1.100; 2.711) & \\
 &\quad Direct effect & 1.821 & (1.095; 2.721) & \\
 &\quad Indirect effect & 1.012 & (0.986; 1.053) & 2.0\%\\
 & \textbf{Death} & & &  \\
 &Total effect ($A=1$ vs. $A=0$) & 0.882 & (0.625; 1.148) & \\
 &\quad Direct effect & 0.882 & (0.617; 1.154) & \\
 &\quad Indirect effect & 1.001 & (0.987; 1.011) & -0.8\%\\
\hline
\multirow{14}{*}{(2)} & \textbf{Dementia} & & &  \\
 &Total effect ($A=1$ vs. $A=0$)  & 1.839 & (1.100; 2.713) & \\
 &\quad  Direct effect & 1.836 & (1.091; 2.716) & \\
 &\quad  Indirect effect & 1.002 & (0.988; 1.017) & 0.3\%\\
 &Total effect ($A=0$ vs. $A=1$)  & 0.544 & (0.369; 0.912) & \\
 &\quad  Direct effect  & 0.554 & (0.367; 0.912) & \\
 &\quad  Indirect effect & 0.981 & (0.925; 1.022) & 3.2\%\\
 & \textbf{Death} & & &  \\
 &Total effect ($A=1$ vs. $A=0$) & 0.882 & (0.622; 1.148) & \\
 &\quad  Direct effect & 0.878 & (0.614; 1.151) & \\
 &\quad  Indirect effect & 1.005 & (0.989; 1.023) & -4.0\%\\
 &Total effect ($A=0$ vs. $A=1$) & 1.134 & (0.871; 1.605) & \\
 &\quad  Direct effect & 1.129 & (0.862; 1.617) & \\
 &\quad Indirect effect & 1.004 & (0.981; 1.036) & 3.2\%\\
\hline
\end{tabular}
\end{table}
As can be seen from table 5, the total effect of hearing loss on the time to dementia diagnosis was statistically significant (Model 1, HR = 1.843; 95\%CI 1.100 to 2.710).  Model (1) further suggested that this total effect was weakly mediated through the feeling of loneliness, with a non-statistically significant indirect effect equal to 1.012 (HR scale; 95\%CI 0.986 to 1.053). This expresses that the hazard of dementia would become 1.012 times higher if all patients were to have limited hearing ability but the loneliness levels were switched from the values that would have been observed if they had normal hearing ability to the value observed under limited hearing. In contrast, the total effect of hearing loss on mortality was not statistically significant  (Model 1, HR = 0.882; 95\%CI 0.625 to 1.148). There was no statistical evidence of an indirect effect through the feeling of loneliness (HR = 1.001, 95\%CI 0.987 to 1.011). These findings did not change when considering model (2) with interaction (table 4 - p-value of the interaction coefficient equals 0.520 for dementia and 0.632 for death).

Figure 3 provides the estimated cumulative incidence curve of dementia for different $a$ and $a^*$, which visualizes the weak indirect effect of hearing loss on dementia through the suggested longitudinal mediator. At some time points, there are large jumps in these curves due to the fairly high rate of interval censoring in the dataset (i.e. if the date for dementia diagnosis (or for death) was not known but person had a new diagnosis (or passed away) from one visit to the next, then we considered the midpoint between two visits as the event date). Finally, the above results should be interpreted with caution as there might be important time-varying confounders of the mediator-outcome association that were not taken into account. The findings could also be biased if the involved models were incorrectly specified or censoring was informative (e.g.  elderly patients who live alone might not come to the control visit due to dementia-related problems).

\section{Discussion}
In this paper, we have generalized the weighting-based strategy proposed for natural effect models in single mediation analysis to the setting where the mediator of interest is repeatedly measured over time (hence subject to longitudinal confounders) and the primary outcome is a time-to-event endpoint, subject to competing risks. The proposed approach yields consistent estimates for the natural direct and indirect effects if the causal assumptions hold and the natural effect model and the conditional distribution of the exposure, mediator and censoring are correctly specified. As noted by Steen et al\citep{steen17.2}, the mediator model needs careful consideration, especially when the exposure (and the baseline covariates) are highly predictive of the mediator, for then even minor misspecification can have a major impact on the weights and lead to biased results. Apart from (mediator) model misspecification, the estimated weights may become unstable or even extreme when at each time point, there is an inadequate overlap between the conditional distributions of the mediator under different treatment/exposure conditions, as may be the case when the exposure has a strong effect on the mediator. While the presence of extreme weights might appear as a limitation at first, it may also diagnose severe model extrapolation that often goes unnoticed when using a repeated regression approach proposed for the same setting \citep{steen17.2}. Simple weighting-based approaches also tend to yield larger standard errors (compared to imputation or regression-based approaches) due to lack in efficiency. This can be especially problematic when the mediator is continuous. In that case, the weight-based approaches tend to be unstable even under proper model specification and adequate overlap of the mediator distributions across treatment groups, which may result in considerable finite sample bias in the natural effect estimates. The repeated regression approach might be more appropriate when dealing with continuous mediators \citep{steen17, vansteelandt2012imputation}.

Several proposals can be made to improve the suggested approach. Future research might focus on the development of doubly or multiply robust estimators \citep{bang2005doubly} to improve the robustness and efficiency of the current weight-based approach. The proposed strategy can also be easily extended to take into account multiple mediators $M^{(1)}, \ldots, M^{(V)}$ that are repeatedly measured over time. When these mediators are causally ordered then as suggested by Vanderweele and Vansteelandt (2014), one can first evaluate the effect mediated through $M^{(1)}$, then examine how much this changes when $M^{(1)}$ and $M^{(2)}$ are jointly considered as mediators. This then reveals the additional contribution of $M^{(2)}$ beyond $M^{(1)}$ alone. The process is then carried on by sequentially adding one mediator at a time until all $V$ mediators are included \citep{vdw14}. By accounting for multiple, repeatedly measured mediators, results of the analysis may allow one to get closer to evaluating the entire mediation process that underlies the treatment mechanism in practice. Finally, future research should also extend the proposed approach to account for continuous exposures, which are quite common in epidemiology and social science.


\section*{Acknowledgments}
The first author was supported by the funding from the European Union’s Horizon 2020 research and innovation program, under the Marie Sklodowska-Curie grant agreement (grant no.: 676207).

\subsection*{Conflict of interest}
The authors declare no potential conflict of interests.

\section*{Supporting information}
The following supporting information is available as part of the online article:

\noindent
\textbf{Data analysis 1.}
{Results of the construction of the propensity score model, the mediator model at each time point and the censoring model}

\noindent
\textbf{Data analysis 2}
{R codes for the analysis of the ELSA data.}

\nocite{*}
\bibliographystyle{wileyNJD-AMA}
\bibliography{longitudinal}%

\appendix
\section{-- Formal proof of the proposal\label{app1}}

We here describe the derivations of the estimating procedure discussed in section 2. Assume that the causal diagram in figure 1  represents a non-parametric structural equation model with independent errors and that the natural effect Cox model (1) is correctly specified. For the sake of simplicity, we first assume that no censoring presents. The added complexity due to censoring will be subsequently addressed. Let $F$ be the score for $(\alpha_1,\alpha_2)$ for all individuals in the sample. To simplify the proof, we will denote $W_i(\lfloor t \rfloor,a,a^*)$ as $W_i(\lfloor t \rfloor)$. One then has:

\begingroup
\fontsize{10pt}{10pt}\selectfont
\begin{align*}
E(F) = \bigintssss \sum_{i, a, a^*} &\left[ \begin{pmatrix} a \\ a^* \end{pmatrix} - \frac{\sum_{a,a^*}\begin{pmatrix} a \\ a^* \end{pmatrix}\cdot e^{\alpha_1a + \alpha_2a^*}\cdot E\left[R_i(t)\cdot W_i(\lfloor t \rfloor)\right]} {\sum_{a,a^*}e^{\alpha_1a + \alpha_2a^*}\cdot E\left[R_i(t)\cdot W_i(\lfloor t \rfloor)\right]}\right]\cdot\\
&\cdot E[R_i(t)\cdot W_i(\lfloor t \rfloor)\cdot dN_i(t)]
\end{align*} 
\endgroup

We now consider the expression $E[R(t)\cdot W(\lfloor t \rfloor)\cdot dN(t)]$, that is: 

\begingroup
\fontsize{10pt}{10pt}\selectfont
\begin{align*}
E[R(t)&\cdot W(\lfloor t \rfloor)\cdot dN(t)] \\
&= E \left\{E(dN(t) | T\ge t, A, \overline{M}_{t}, \overline{L}_{t}, L_{0})\cdot R(t)\cdot W(\lfloor t \rfloor) \right\}\\
&=E \left\{E(dN(t) | T\ge t, A, \overline{M}_{t}, \overline{L}_{t}, L_{0}) \cdot R(t) \cdot W^*(\lfloor t \rfloor)\,\bigg\rvert\,A=a \right\}\pr(A=a)
\end{align*}
\endgroup

where 

\begingroup
\fontsize{10pt}{10pt}\selectfont
\[W^*(\lfloor t \rfloor) =  \frac{\prod_{s:t_s\le \lfloor t \rfloor} \pr(M_{s}|A = a^*, \overline{M}_{s - 1}, \overline{L}_{s}, T\ge t_s)}{\prod_{s:t_s\le \lfloor t \rfloor} \pr(M_{s}|A = a, \overline{M}_{s - 1}, \overline{L}_{s}, T\ge t_s)}\times \frac{1}{\pr(A=a|L_{0})}.\] 
\endgroup

Denote $t-=t - \delta t$ where $\delta t > 0$ is a small positive quantity such that $\lfloor t \rfloor < t-<t$. One then has:

\begingroup
\fontsize{10pt}{10pt}\selectfont
\begin{align*}
E[&R(t)\cdot W(t)\cdot dN(t)] \\
&= E \bigg\{E(dN(t)| T\ge t, a, \overline{M}_{\lfloor t \rfloor},\overline{L}_{\lfloor t \rfloor}, L_{0})\cdot I(T\ge t)\cdot I(T\ge t-)\cdot W^*(\lfloor t \rfloor)\,\bigg\rvert\, A=a \bigg\}\pr(a)\\
&= E \bigg\{E\left[E(dN(t) | T\ge t, a, \overline{M}_{\lfloor t \rfloor},\overline{L}_{\lfloor t \rfloor}, L_{0})\cdot I(T\ge t)\cdot I(T\ge t-)\cdot\right.\\
&~~~~~~~~~~ \left.\cdot W^*(\lfloor t \rfloor)\,\bigg\rvert\, T\ge t-,A=a,M_{\lfloor t \rfloor},L_{\lfloor t \rfloor},L_0\right]\,\bigg\rvert\, A=a \bigg\}\pr(a)\\
&=E \bigg\{E(dN(t) | T\ge t, a,\overline{M}_{\lfloor t \rfloor},\overline{L}_{\lfloor t \rfloor}, L_{0})\cdot I(T\ge t-)\cdot W^*(\lfloor t \rfloor)\cdot\\
&~~~~~~~~~~ \cdot \pr\bigg[T\ge t|T\ge t-,a,\overline{M}_{\lfloor t \rfloor},\overline{L}_{\lfloor t \rfloor},L_0\bigg] \,\bigg\rvert\, A=a\bigg\}\pr(a)
\end{align*}
\endgroup

where $E(dN(t)| T\ge t, a, \overline{M}_{\lfloor t \rfloor},\overline{L}_{\lfloor t \rfloor}, L_{0})$ is a short-hand notation for $E(dN(t)| T\ge t, A=a, \overline{M}_{\lfloor t \rfloor},\overline{L}_{\lfloor t \rfloor}, L_{0})$ and so on. Note that $\pr\left[T\ge t|T\ge t-,A=a,\overline{M}_{\lfloor t \rfloor},\overline{L}_{\lfloor t \rfloor},L_0\right] = 1 - \lambda(t|a,\overline{M}_{\lfloor t \rfloor},\overline{L}_{\lfloor t \rfloor},L_0)\delta t$. This implies that:

\begingroup
\fontsize{10pt}{10pt}\selectfont
\begin{align*}
E[&R(t)\cdot W(t)\cdot dN(t)] =E \bigg\{E(dN(t) | T\ge t, a,\overline{M}_{\lfloor t \rfloor},\overline{L}_{\lfloor t \rfloor}, L_{0})\cdot I(T\ge t-)\cdot \\
&~~~~~~~~~~~~~~~~~~~~~~~~~~~~~~~~~~~~~~~~\cdot W^*(\lfloor t \rfloor)\cdot \left[ 1 - \lambda(t|a,\overline{M}_{\lfloor t \rfloor},\overline{L}_{\lfloor t \rfloor},L_0)\delta t\right] \,\bigg\rvert\,A=a \bigg\}\pr(a)
\end{align*}
\endgroup

Applying the above arrangements in a backward fashion till when the time point $\lfloor t \rfloor$ is reached, one then obtains:

\begingroup
\fontsize{10pt}{10pt}\selectfont
\begin{align*}
E[&R(t)\cdot W(t)\cdot dN(t)] \\
&=E \bigg\{E(dN(t) | T\ge t, a,\overline{M}_{\lfloor t \rfloor},\overline{L}_{\lfloor t \rfloor}, L_{0})\cdot I(T\ge \lfloor t\rfloor)\cdot W^*(\lfloor t \rfloor)\cdot \\
& ~~~~~~~~~~\bm\prod_{s: \lfloor t \rfloor \le s \le t} \left[ 1 - \lambda(s|a,\overline{M}_{\lfloor t \rfloor},\overline{L}_{\lfloor t \rfloor},L_0)\right] \,\bigg\rvert\,A=a  \bigg\}\pr(a)\\
&=E\left\{\bigintssss E (dN(t) | T\ge t, a,m_{\lfloor t \rfloor},\overline{M}_{\lfloor t-1 \rfloor},l_{\lfloor t \rfloor},\overline{L}_{\lfloor t-1 \rfloor}, L_{0})\cdot I(T\ge \lfloor t\rfloor)\cdot W^*(\lfloor t-1 \rfloor)\,\cdot \right.\\
&\,\,\,\,\,\,\,\,\,\,\,\,\,\,\left.\cdot \,\pr(m_t|a^*,\overline{M}_{\lfloor t-1 \rfloor},\overline{L}_{\lfloor t \rfloor},L_0,T\ge t)\cdot \pr(l_t|a,\overline{L}_{\lfloor t-1 \rfloor},\overline{M}_{\lfloor t-1 \rfloor},L_0,T\ge t)\,\cdot \right.\\
&\,\,\,\,\,\,\,\,\,\,\,\,\,\,\left.\cdot \bm\prod_{s: \lfloor t \rfloor \le s \le t} \left[ 1 - \lambda(s|a,m_{\lfloor t \rfloor},\overline{M}_{\lfloor t-1 \rfloor},l_{\lfloor t \rfloor},\overline{L}_{\lfloor t-1 \rfloor},L_0)\right] \,dm_{\lfloor t \rfloor}\,dl_{\lfloor t \rfloor}\vphantom{\frac{a}{b}}\,\bigg\rvert\,A=a \right\}\pr(a)
\end{align*}
\endgroup

where $\bm\prod_{s} a_s$ is defined as a product limit. The above procedure continues to be repeated in a backward fashion till when the starting time point $t = 0$ is reached, which then gives:

\begingroup
\fontsize{10pt}{10pt}\selectfont
\begin{align*}
E[&R(t)\cdot W(t)\cdot dN(t)] \\
&=E\bigg\{\bigintssss  E(dN(t) | T\ge t, a,\overline{m}_{\lfloor t \rfloor}, \overline{l}_{\lfloor t \rfloor}, L_{0}) \cdot \bm \prod_{s: 0<s \le t} \left[ 1 - \lambda(s|a,\overline{m}_{\lfloor s \rfloor},\overline{l}_{\lfloor s \rfloor},L_0)\right]\cdot \\
&\,\,\,\,\,\,\,\,\,\,\,\,\,\, \cdot \prod_{s: 0<s\le \lfloor t \rfloor} \pr(m_{s}| a^*, \overline{m}_{s - 1}, \overline{l}_{s}, L_{0}, T\ge t_s) \cdot\\
&~~~~~~~~ \cdot \pr(l_{s}|\overline{l}_{s-1}, \overline{m}_{s - 1}, L_0, a, T\ge t_s)\,d\overline{m}_{\lfloor t \rfloor}\,d\overline{l}_{\lfloor t \rfloor}\,\bigg\rvert\,A=a \bigg\}\pr(a)
\end{align*}
\endgroup

Note that $ \bm \prod_{s: 0<s \le t} \left[ 1 - \lambda(s|a,\overline{m}_{\lfloor s \rfloor},\overline{l}_{\lfloor s \rfloor},L_0)\right] = \pr(T\ge t|a,\overline{m}_{\lfloor t \rfloor},\overline{l}_{\lfloor t \rfloor},L_0)$. This implies that:

\begingroup
\fontsize{10pt}{10pt}\selectfont
\begin{align*}
E[&R(t)\cdot W(t)\cdot dN(t)] \\
&=E\bigg\{\bigintssss  E (dN(t) | T\ge t, a,\overline{m}_{\lfloor t \rfloor}, \overline{l}_{\lfloor t \rfloor}, L_{0}) \cdot \pr(T\ge t|a,\overline{m}_{\lfloor t \rfloor},\overline{l}_{\lfloor t \rfloor},L_0)\cdot \\
&\,\,\,\,\,\,\,\,\,\,\,\,\,\, \cdot \prod_{s: 0<s\le t} \pr(m_{s}| a^*, \overline{m}_{s - 1}, \overline{l}_{s}, L_{0}, T\ge t_s) \cdot \\
&\,\,\,\,\,\,\,\,\,\,\,\,\,\, \cdot \pr(l_{s}|\overline{l}_{s-1}, \overline{m}_{s - 1}, L_0, a, T\ge t_s),d\overline{m}_{\lfloor t \rfloor}\,d\overline{l}_{\lfloor t \rfloor}\,\bigg\rvert\,A=a \bigg\}\pr(a)\\
&= \bigintssss  E( dN(t)|a, \overline{m}_{t}, \overline{l}_{t}, l_{0})\cdot \prod_{s:0<s\le\lfloor t \rfloor} \pr(m_{s}| a^*, \overline{m}_{s - 1}, \overline{l}_{s}, l_{0}, T\ge t_s)\cdot \\
&\,\,\,\,\,\,\,\,\,\,\,\,\,\,\cdot \pr(l_{s}|\overline{l}_{s-1}, \overline{m}_{s - 1}, l_{0}, a, T\ge t_s)\cdot \frac{\pr(l_{0}|a)\times \pr(a)}{\pr(a|l_0)}\,d\overline{m}_{\lfloor t \rfloor}\,d\overline{l}_{\lfloor t \rfloor}\,d l_0\\
&= E\left[dN^{a,a^*}(t)\right] \\
&=  E\left[R^{a,a^*}(t)\right]\cdot  E\left[dN^{a,a^*}(t)|R^{a,a^*}(t)=1\right]
\end{align*}
\endgroup

where the prior-to-last equaling follows by the derivation of Vansteelandt et al\citep{vansteelandt19}. In a similar way, one can show that $
E[R(t) \cdot W(t)]=  E \left(R^{a, a^*}(t)\right)$. These imply that:

\begingroup
\fontsize{10pt}{10pt}\selectfont
\begin{align*}
E(F) &=\bigintsss \sum_{i, a, a^*} \left[ \begin{pmatrix} a \\ a^* \end{pmatrix} - \frac{\sum_{a,a^*}\begin{pmatrix} a \\ a^* \end{pmatrix}\cdot e^{\alpha_1a + \alpha_2a^*}\cdot E\left[R^{a,a^*}_i(t)\right]} {\sum_{a,a^*}e^{\alpha_1a + \alpha_2a^*}\cdot E\left[R^{a,a^*}_i(t)\right]}\right]\cdot \\
&\,\,\,\,\,\,\,\,\,\,\,\,\,\,\,\,\,\,\, \cdot  E\left[R^{a,a^*}(t)\right]\cdot  E\left[dN^{a,a^*}(t)|R^{a,a^*}(t)=1\right]\\
&=\bigintsss \sum_{i, a, a^*} \left[ \begin{pmatrix} a \\ a^* \end{pmatrix} - \frac{\sum_{a,a^*}\begin{pmatrix} a \\ a^* \end{pmatrix}e^{\alpha_1a + \alpha_2a^*}\cdot E\left[R^{a,a^*}_i(t)\right]} {\sum_{a,a^*}e^{\alpha_1a + \alpha_2a^*}\cdot E\left[R^{a,a^*}_i(t)\right]}\right]\cdot \\
&\,\,\,\,\,\,\,\,\,\,\,\,\,\,\,\,\,\,\, \cdot E\left[R^{a,a^*}(t)\right]\cdot \left\{ E\left[dN^{a,a^*}(t)|R^{a,a^*}(t)=1\right] - \lambda_0(t)\cdot e^{\alpha_1a+\alpha_2a^*}\right\}\\
&=0
\end{align*}
\endgroup

The final equaling follows by the fact that the natural proportional hazard model is correctly specified, for then $E[dN_i(t)^{a, a^*} (t)|T_i^{a, a^*} \ge t] = \lambda_0(t)\cdot e^{\alpha_1a+\alpha_2a^*}$
\section{-- Addressing informative censoring} 
When censoring presents but is non-informative given the exposure group, the score $F$ for $(\alpha_1, \alpha_2)$ for all individuals in the sample can be written as:

\begingroup
\fontsize{10pt}{10pt}\selectfont
 \begin{align*}
F = &\bigintssss_0^\infty \sum_{i, a, a^*}\left\{\begin{pmatrix} a \\ a^* \end{pmatrix} - \frac{\sum_{a,a^*}E\left[\begin{pmatrix} a \\ a^* \end{pmatrix}\cdot R_i(t)\cdot I(C_i>t)\cdot W_i(\lfloor t \rfloor )\cdot e^{\alpha_1a + \alpha_2a^*}\right]}
{\sum_{a,a^*} E\left[R_i(t)\cdot I(C_i>t)\cdot W_i(\lfloor t \rfloor )\cdot e^{\alpha_1a + \alpha_2a^*}\right]}\right\}\cdot \\
&\,\,\,\,\,\,\,\,\,\,\,\,\,\,\,\,\,\,\,\,\,\,\,\,\,\,\,\,\,\,\cdot R_i(t)\cdot I(C_i>t)\cdot W_i(\lfloor t \rfloor )\left(dN_i(t) - \lambda_0(t)\cdot e^{\alpha_1a + \alpha_2a^*}dt\right)
\end{align*}
\endgroup

Note that:

\begingroup
\fontsize{10pt}{10pt}\selectfont
\begin{align*}
E[&R(t)\cdot I(C\ge t) \cdot W(t)\cdot dN(t)] \\
 &= E \left\{ E\left[I(C\ge t)\cdot R(t) \cdot dN(t) \cdot W(\lfloor t \rfloor) \,\bigg\rvert\, A,\overline{M}_{\lfloor t\rfloor }, \overline{L}_{\lfloor t\rfloor }, L_{0} \right]\right\}\\
&= E \bigg\{ W(\lfloor t \rfloor)\cdot \pr(C\ge t|A,\overline{M}_{\lfloor t\rfloor }, \overline{L}_{\lfloor t\rfloor }, L_0, T\ge t) \cdot \\
&~~~~\cdot E[dN(t)|A,\overline{M}_{\lfloor t\rfloor }, \overline{L}_{\lfloor t\rfloor}, L_0, T\ge t]
\cdot \pr(T\ge t|A,\overline{M}_{\lfloor t\rfloor }, \overline{L}_{\lfloor t\rfloor }, L_0)\bigg\}\\
&= E \left\{ W(\lfloor t \rfloor)\cdot \pr(C\ge t|A)
\cdot E[dN(t)|A,\overline{M}_{\lfloor t\rfloor }, \overline{L}_{\lfloor t\rfloor}, L_0, T\ge t]
\cdot \pr(T\ge t|A,\overline{M}_{\lfloor t\rfloor }, \overline{L}_{\lfloor t\rfloor }, L_0)\right\}\\
&= \pr(C\ge t|A=a)\cdot E \bigg\{ W^*(\lfloor t \rfloor)
\cdot E[dN(t)|a,\overline{M}_{\lfloor t\rfloor }, \overline{L}_{\lfloor t\rfloor}, L_0, T\ge t]\\
&~~~~\cdot \pr(T\ge t|a,\overline{M}_{\lfloor t\rfloor }, \overline{L}_{\lfloor t\rfloor }, L_0) \,\bigg\rvert\,A=a\bigg\}\pr(a)
\end{align*}
\endgroup

The prior-to-last equaling results from the assumption that the censoring is non-informative given the treatment group (e.g. figure 3). Following the same reasoning as above, one can show that:

\begin{align*}
E[R(t)\cdot I(C\ge t) \cdot W(t)\cdot dN(t)] = \pr(C\ge t|A=a)\cdot E\left[R^{a,a^*}(t)\right]\cdot \\
\cdot E\left[dN^{a,a^*}(t)|R^{a,a^*}(t)=1\right]~~~~~~
\end{align*}
and similarly,
\begin{align*}
E[R(t)\cdot I(C\ge t) \cdot W(t)] = \pr(C\ge t|A=a)\cdot E\left[R^{a,a^*}(t)\right]
\end{align*}
The score will thus remain the same as in the case of no censoring, with an added proportional constant, i.e. $ \pr(C>t|A=a)$. As a result,

\begingroup
\fontsize{10pt}{10pt}\selectfont
\begin{align*}
E(F)&=\bigintssss \sum_{i, a, a^*}  E\left\{\left[\begin{pmatrix} a \\ a^* \end{pmatrix} - \frac{\sum_{a,a^*}\begin{pmatrix} a \\ a^* \end{pmatrix}\cdot e^{\alpha_1a + \alpha_2a^*}\cdot E\left(R_i^{a, a^*}(t)\right)\cdot \pr(C>t|A=a)}{\sum_{a,a^*}e^{\alpha_1a + \alpha_2a^*}\cdot E\left(R_i^{a, a^*}(t)\right)\cdot \pr(C>t|A=a)}\right]\cdot\right.\\
&\,\,\,\,\,\,\,\left.\cdot\, E\left[R_i^{a, a^*}(t)\right]\cdot \left[E[dN_i(t)^{a, a^*} (t)|T_i^{a, a^*} \ge t] - \lambda_0(t)\cdot e^{\alpha_1a+\alpha_2a^*}\right] \right\}\cdot\pr(C>t|A=a)\\
&= 0
\end{align*}
\endgroup

which results from the fact that the natural proportional hazard model is correctly specified. 

When censoring is only non-informative at time $t$ given the exposure group and the history of mediator and covariates up to that time, the weight of each patient at time $t$ needed to be adjusted as described in section 2.3. To prove this, we consider once again the expression $E[R(t)\cdot I(C>t)\cdot W(\lfloor t \rfloor) \cdot dN(t)]$. Note that the weight $W(\lfloor t \rfloor)$ now has an additional censoring-weight component (see section 2.3), for then:

\begingroup
\fontsize{10pt}{10pt}\selectfont
\begin{align*}
E[&R(t)\cdot I(C>t)\cdot W(\lfloor t \rfloor) \cdot dN(t)]\\
&= E \left\{E(dN(t) | C\ge t, T\ge t, A, \overline{M}_{\lfloor t \rfloor}, \overline{L}_{\lfloor t \rfloor}, L_{0})\cdot R(t)\cdot I(C>t) \cdot W(\lfloor t \rfloor) \right\}\\
&=E \left\{E(dN(t) | T\ge t, A, \overline{M}_{\lfloor t\rfloor }, \overline{L}_{\lfloor t \rfloor }, L_{0}) \cdot R(t) \cdot I(C>t) \cdot W^*(\lfloor t \rfloor)\,\bigg\rvert\,A=a \right\}\pr(A=a)
\end{align*}
\endgroup

where 

\begingroup
\fontsize{10pt}{10pt}\selectfont
\begin{align*}
W^*(t) =& \frac{\prod_{s: 0\le s \le \lfloor t \rfloor} \pr(M_{s}|A = a^*, \overline{M}_{\lfloor s \rfloor-1}, \overline{L}_{\lfloor s \rfloor}, T\ge t_s, C \ge t_s)}{\prod_{s: 0\le s \le \lfloor t \rfloor} \pr(M_{s}|A = a, \overline{M}_{\lfloor s\rfloor-1}, \overline{L}_{\lfloor s \rfloor}, T\ge t_s, C \ge t_s)}\cdot \frac{1}{\pr(A=a|L_{0})}\cdot\\
&\,\cdot\frac{1}{\prod_{s: 0\le s \le t} \left[1- \lambda_C(s|T>s, \overline{m}_{s }, \overline{l}_{s},l_{0},a)\right]}
\end{align*}
\endgroup

Note that the second equaling results from the fact that censoring is non-informative at any time $t$, given the history up to that time. Now denote $t-=t - \delta t$ where $\delta t > 0$ is a small positive quantity such that $\lfloor t \rfloor < t-<t$, one then has:

\begingroup
\fontsize{10pt}{10pt}\selectfont
\begin{align*}
E&[R(t)\cdot I(C>t)\cdot W(\lfloor t \rfloor) \cdot dN(t)]\\
&= E \bigg\{E(dN(t)| T\ge t, a, \overline{M}_{\lfloor t \rfloor},\overline{L}_{\lfloor t \rfloor}, L_{0})\cdot I(T\ge t)\cdot I(T\ge t-)\cdot\\
&\quad \quad\cdot I(C\ge t) \cdot I(C\ge t-) \cdot W^*(\lfloor t \rfloor)\,\bigg\rvert\,A=a \bigg\}\pr(a)\\
&= E \biggl\{E\biggl[E(dN(t) | T\ge t, a, \overline{M}_{\lfloor t \rfloor},\overline{L}_{\lfloor t \rfloor}, L_{0})\cdot I(T\ge t)\cdot I(T\ge t-)\cdot I(C\ge t)\cdot  \nonumber\\
&\quad\quad \cdot I(C\ge t-)\cdot W^*(\lfloor t \rfloor)\,\bigg\rvert\,T\ge t-, C\ge t-, A=a,M_{\lfloor t \rfloor},L_{\lfloor t \rfloor},L_0\biggr]\,\bigg\rvert\, A=a \biggr\}\pr(a)\\
&=E \biggl\{E(dN(t) | T\ge t, a,\overline{M}_{\lfloor t \rfloor},\overline{L}_{\lfloor t \rfloor}, L_{0})\cdot I(T\ge t-)\cdot I(C\ge t-) \cdot W^*(\lfloor t \rfloor)\cdot \nonumber\\
&\quad \quad \cdot \Esp\biggl[I(T\ge t)\cdot I(C\ge t)\,\bigg\rvert\,T\ge t-,C\ge t-,a,\overline{M}_{\lfloor t \rfloor},\overline{L}_{\lfloor t \rfloor},L_0\biggr] \,\bigg\rvert\, A=a\biggr\}\pr(a)\\
&=E \biggl\{E(dN(t) | T\ge t, a,\overline{M}_{\lfloor t \rfloor},\overline{L}_{\lfloor t \rfloor}, L_{0})\cdot I(T\ge t-)\cdot I(C\ge t-) \cdot W^*(\lfloor t \rfloor)\nonumber\\
&\quad\quad \cdot \pr[C\ge t|T\ge t,C\ge t-,a,\overline{M}_{\lfloor t \rfloor},\overline{L}_{\lfloor t \rfloor},L_0]\cdot \\
&\quad \quad\cdot \pr[T\ge t|T\ge t-,C\ge t-,a,\overline{M}_{\lfloor t \rfloor},\overline{L}_{\lfloor t \rfloor},L_0] \,\bigg\rvert\, A=a\biggr\}\pr(a)\\
&=E \biggl\{E(dN(t) | T\ge t, a,\overline{M}_{\lfloor t \rfloor},\overline{L}_{\lfloor t \rfloor}, L_{0})\cdot I(T\ge t-)\cdot I(C\ge t-) \cdot W^*(\lfloor t \rfloor)\nonumber\\
&\quad\quad \cdot [1-\lambda_C(t|T\ge t,a,\overline{M}_{\lfloor t \rfloor},\overline{L}_{\lfloor t \rfloor},L_0)\delta t]\cdot \\
&\quad \quad\cdot \pr[T\ge t|T\ge t-,a,\overline{M}_{\lfloor t \rfloor},\overline{L}_{\lfloor t \rfloor},L_0] \,\bigg\rvert\, A=a\biggr\}\pr(a)\\
&=E \biggl\{E(dN(t) | T\ge t, a,\overline{M}_{\lfloor t \rfloor},\overline{L}_{\lfloor t \rfloor}, L_{0})\cdot I(T\ge t-)\cdot I(C\ge t-) \cdot W^*(\lfloor t \rfloor)\nonumber\\
&\quad\quad \cdot [1-\lambda_C(t|T\ge t,a,\overline{M}_{\lfloor t \rfloor},\overline{L}_{\lfloor t \rfloor},L_0)\delta t]\cdot [1-\lambda(t|a,\overline{M}_{\lfloor t \rfloor},\overline{L}_{\lfloor t \rfloor},L_0)\delta t] \,\bigg\rvert\, A=a\biggr\}\pr(a)
\end{align*}
\endgroup

The last two equalings result from the fact that $\pr\left[C\ge t|T\ge t,C\ge t-,A=a,\overline{M}_{\lfloor t \rfloor},\overline{L}_{\lfloor t \rfloor},L_0\right] = 1 - \lambda_C(t|T\ge t, a,\overline{M}_{\lfloor t \rfloor},\overline{L}_{\lfloor t \rfloor},L_0)\delta t$ and that $\pr[T\ge t|T\ge t-,a,\overline{M}_{\lfloor t \rfloor},\overline{L}_{\lfloor t \rfloor},L_0] =  \pr[T\ge t|T\ge t-,C\ge t-,a,\overline{M}_{\lfloor t \rfloor},\overline{L}_{\lfloor t \rfloor},L_0]$ due to the assumption of conditionally non-informative censoring. This implies that:

\begingroup
\fontsize{10pt}{10pt}\selectfont
\begin{align*}
E[&R(t)\cdot I(C>t)\cdot W(\lfloor t \rfloor) \cdot dN(t)]\\
&=E \biggl\{E(dN(t) | T\ge t, a,\overline{M}_{\lfloor t \rfloor},\overline{L}_{\lfloor t \rfloor}, L_{0})\cdot I(T\ge t-)\cdot I(C\ge t-)\, \cdot \nonumber\\
&\quad\quad\cdot \frac{\prod_{s: 0\le s \le \lfloor t \rfloor} \pr(M_{s}|A = a^*, \overline{M}_{\lfloor s \rfloor-1}, \overline{L}_{\lfloor t \rfloor}, T\ge t_s, C \ge t_s)}{\prod_{s: 0\le s \le \lfloor t \rfloor} \pr(M_{s}|A = a, \overline{M}_{\lfloor s\rfloor-1}, \overline{L}_{\lfloor s \rfloor}, T\ge t_s, C \ge t_s)}\cdot \frac{1}{\pr(A=a|L_{0})}\cdot\\
&\quad\quad\cdot \frac{1-\lambda(t|a,\overline{M}_{\lfloor t \rfloor},\overline{L}_{\lfloor t \rfloor},L_0)}{\prod_{s: 0\le s \le t-} \left[1- \lambda_C(s|T>s, \overline{m}_{s }, \overline{l}_{s},l_{0},a)\right]} \,\bigg\rvert\, A=a\biggr\}\pr(a)
\end{align*}
\endgroup

Applying the above arrangements in a backward fashion as is done in the case of no censoring, one will obtain the result that:

\begingroup
\fontsize{10pt}{10pt}\selectfont
\[E[R(t)\cdot I(C>t)\cdot W(\lfloor t \rfloor) \cdot dN(t)]=E\left[R^{a,a^*}(t)\right]\cdot  E\left[dN^{a,a^*}(t)|R^{a,a^*}(t)=1\right]\] 
\endgroup

The rest of the proof is thus the same as in the case of no censoring.

\clearpage

\end{document}